\documentclass[twocolumn,times]{aastex63} 
\usepackage{amsmath,bm}
\usepackage{xcolor}
\usepackage[english]{babel}
\usepackage[latin1]{inputenc}
\usepackage{indentfirst}
\usepackage{booktabs}
\usepackage{multirow}

\usepackage{threeparttable}

\newcommand{\mpchi}{\,h^{-1}{\rm {Mpc}}}

\newcommand{\msun}{M_{\sun}}
\newcommand{\msunh}{h^{-1}M_{\sun}}

\newcommand{\hi}{H{~\sc i}}

\defcitealias{Guo2017}{G17}
\defcitealias{Guo2020}{G20}

\received{XXX}
\revised{XXX}
\accepted{XXX}
\submitjournal{ApJ}

\shorttitle{\hi-halo Mass Relation}
\shortauthors{Li, Guo \& Mao}

\begin{document}

\title{Theoretical Models of the Atomic Hydrogen Content in Dark Matter Halos}

\correspondingauthor{Hong Guo, Yi Mao}
\email{guohong@shao.ac.cn (HG), ymao@tsinghua.edu.cn (YM)}

\author[0000-0002-1287-6064]{Zhixing Li}
\affiliation{Department of Astronomy, Tsinghua University, Beijing 100084, China}

\author[0000-0003-4936-8247]{Hong Guo}
\affiliation{Shanghai Astronomical Observatory, Chinese Academy of Sciences, Shanghai 200030, China}
\author[0000-0002-1301-3893]{Yi Mao}
\affiliation{Department of Astronomy, Tsinghua University, Beijing 100084, China}

\begin{abstract}
    Atomic hydrogen (H{~\sc i}) gas, mostly residing in dark matter halos after cosmic reionization, is the fuel for star formation. Its relation with properties of host halo is the key to understand the cosmic H{~\sc i} distribution. In this work, we propose a flexible, empirical model of H{~\sc i}-halo relation. In this model, while the H{~\sc i} mass depends primarily on the mass of host halo, there is also secondary dependence on other halo properties. We apply our model to the observation data of the Arecibo Fast Legacy ALFA Survey (ALFALFA), and find it can successfully fit to the cosmic H{~\sc i} abundance ($\Omega_{\rm HI}$), average H{~\sc i}-halo mass relation $\langle M_{\rm HI}|M_{\rm h}\rangle$, and the H{~\sc i} clustering. The bestfit of the ALFALFA data rejects with high confidence level the model with no secondary halo dependence of H{~\sc i} mass and the model with secondary dependence on halo spin parameter ($\lambda$), and shows strong dependence on halo formation time ($a_{1/2}$) and halo concentration ($c_{\rm vir}$). In attempt to explain these findings from the perspective of hydrodynamical simulations, the IllustrisTNG simulation confirms the dependence of H{~\sc i} mass on secondary halo parameters. However, the IllustrisTNG results show strong dependence on $\lambda$ and weak dependence on $c_{\rm vir}$ and $a_{1/2}$, and also predict a much larger value of H{~\sc i} clustering on large scales than observations. This discrepancy between the simulation and observation calls for improvements in understanding the H{~\sc i}-halo relation from both theoretical and observational sides.
\end{abstract}


\section{Introduction}
Atomic hydrogen (\hi) gas is the fuel for early formation of stars and galaxies. The \hi\ gas is harbored mainly in the cold and dense regions of dark matter halos after cosmic reionization, where it is self-shielded from the UV background with high recombination rates \citep[see e.g.,][]{Krumholz2009, Diemer2018}, making \hi\ emission a biased tracer of the dark matter distribution in the universe \citep{Zhenyuan2021}.

The 21~cm radiation due to the hyperfine transition of atomic hydrogen is a novel probe of the large-scale structure of the universe \citep{Abdalla2005, Furlanetto2006}. Intensity mapping technology has been proposed and applied in many 21~cm surveys as a powerful tool to probe the large scale structure without resolving individual sources in modest 3D pixels, which implies a shorter integration time and lower angular resolution to get reasonable statistics on large scale \citep[see e.g.,][]{Bull2015}. Based on this technology, some pioneering experimental 21~cm intensity mapping surveys have been conducted \citep{Masui2013, Switzer2013}. On the basis of these preceding attempts, the next generation telescopes and surveys have been proposed or already under construction to detect the post-reionization universe, such as the BAO from Integrated Neutral Gas Observations \citep[BINGO;][]{Battye2012}, Tianlai \citep{Xuelei2012, Yidong2014}, Canadian Hydrogen Intensity Mapping Experiment \citep[CHIME;][]{Newburgh2014} and the Square Kilometre Array \citep[SKA;][]{Dewdney2009,Bacon2020}.
     
One key element to understand the large-scale structure using the statistics of the 21~cm observations, especially the \hi\ power spectrum, is the relation between the \hi\ gas and the dark matter halo masses. In principle, the \hi\ gas mass within and around a given galaxy is affected by the star formation and various feedback processes in the galaxy baryon cycle \citep{Tumlinson2017}. In reality, considering the total \hi\ mass in a dark matter halo, it is still largely determined by the halo mass as found by recent hydrodynamic simulations and semi-analytical models (SAMs) \citep{Villaescusa2018,Baugh2019,Chauhan2020,Spinelli2020}. With an accurate \hi-halo mass relation measured and modeled, it can be used to understand the evolution of the \hi\ with the environment, as well as the large-scale structure. Mock catalogs for the future surveys could also be conveniently constructed by applying this relation to the cosmological simulations \citep{Bagla2010, Villaescusa2014, Seehars2016}. 

However, hydrodynamic simulations and SAMs employ different galaxy formation models, and thus predict different \hi-halo mass relations \citep[see e.g. Figure~8 of][]{Baugh2019}. Various empirical laws have also been proposed to describe the \hi-halo mass relation \citep[see e.g.,][]{Villaescusa2018,Baugh2019,Obuljen2019,Spinelli2020}, but they are typically designed to fit to specific simulations or SAMs. It is thus important to measure the \hi-halo mass relation in observations, which, however, is challenging, because the individual measurements of the \hi\ mass in galaxies can suffer from the effect of flux limits of the 21~cm surveys, i.e.\ only \hi-rich galaxies can be observed \citep{Haynes2011}. 

Recently, \cite{Guo2020} (hereafter \citetalias{Guo2020}) presented a direct observational measurement of the \hi-halo mass relation at $z\simeq 0$ by stacking the overall \hi\ signals from the Arecibo Fast Legacy ALFA Survey \citep[ALFALFA;][]{Giovanelli2005} for halos of different masses, constructed from the galaxy group catalog \citep{Lim2017}. They found that the \hi\ mass is not a simple monotonically increasing function of halo mass, but shows strong, additional dependence on the halo richness. That is consistent with the finding of \cite{Guo2017} (hereafter \citetalias{Guo2017}) that \hi-rich galaxies tend to live in halos with late-formation time, using the spatial clustering measurements, as confirmed by \cite{Stiskalek2021}. Such a halo assembly bias effect \citep[see e.g.,][]{Gao2005,Jing2007} in the \hi-halo mass relation is generally not taken into account in most of the previous empirical models. 
     
In this paper, we extend the work of \citetalias{Guo2020} by proposing a more flexible, empirical model for the \hi-halo relation at $z\simeq 0$, which includes the halo assembly bias effect. With the information of the \hi\ abundance from the \hi-halo mass relation and the \hi\ bias from the spatial clustering of \hi-selected galaxies, we aim to obtain an accurate theoretical model for the \hi-halo mass relation, which can be extended to the range of smaller halo mass than what is probed observationally. Also, the dependence on the halo assembly history encodes information about the evolution of the total \hi\ gas with the halo mass, which can potentially provide forecasts for the future \hi\ surveys at redshifts higher than $z\simeq 0$.  
     
The rest of this paper is organized as follows. We introduce the observational measurements and our empirical model in Section~\ref{sec:observation methodology}, and present the results of fitting our model to the observation data in Section~\ref{sec:observational result}. These results are compared with the hydrodynamic simulations and the previous works in Section~\ref{sec:simulation part}. We make concluding remarks in Section~\ref{sec:conclusion}.
     
\section{Measurements and Methodology}
\label{sec:observation methodology}

\subsection{Observational Measurements}\label{sec:data}
We use three sets of measurements to constrain the \hi-halo mass relation --- the cosmic \hi\ abundance ($\Omega_{\rm HI}$), the average \hi-halo mass relation ($\langle M_{\rm HI}|M_{\rm h}\rangle$), and the \hi\ clustering measurements ($w_{\rm p}(r_{\rm p})$ and $\xi(s)$).
  
\subsubsection{Cosmic H{\scriptsize I} Abundance}     
The dimensionless cosmic \hi\ abundance, $\Omega_{\rm HI}$, can be calculated as
    \begin{equation}
        \Omega_{\rm HI}=\frac{1}{\rho_{\rm c}}\int \langle M_{\rm HI}|M_{\rm h}\rangle n(M_{\rm h})dM_{\rm h}, \label{eq:omegahi}
    \end{equation}
where $n(M_{\rm h})$ is the halo mass function, and $\rho_{\rm c}$ is the critical density. It describes the total amount of \hi\ in the universe. We adopt the value of $\Omega_{\rm HI}=(3.5\pm0.6)\times10^{-4}$ from \cite{Jones2018}. It was obtained from the observed \hi\ mass function of ALFALFA 100\% sample \citep{Haynes2018} by integrating the assumed Schechter function. 

The measurement of $\Omega_{\rm HI}$ is used to constrain the low mass part of the \hi-halo mass relation. We note that it is the original value before applying the correction for \hi\ self-absorption, to be consistent with the other two measurements of $\langle M_{\rm HI}|M_{\rm h}\rangle$ and observed \hi\ masses of individual galaxies in ALFALFA, where self-absorption is also not corrected. The \hi\ self-absorption is difficult to be accurately estimated in observations and thus controversial.

\subsubsection{\rm{H{\scriptsize I}}-halo Mass Relation}
We use the measured \hi-halo mass relation $\langle M_{\rm HI}|M_{\rm h}\rangle$ of \citetalias{Guo2020} in the halo mass range of $10^{11}$--$10^{14}\msunh$, with a bin size of $\Delta\log M_{\rm h}=0.25$ (shown as open circles in the left panel of Figure~\ref{fig:obs}). To fit the observational measurements with a finite bin size, the average \hi\ mass in halo mass bins of $M_{\rm h1}<M_{\rm h}<M_{\rm h2}$ should be modeled as,
    \begin{equation}
    \label{eq:avg}
        \langle M_{\rm HI}|M_{\rm h}\rangle_{\rm obs}\big|^{M_{\rm h2}}_{M_{\rm h1}}=\dfrac{\int_{M_{\rm h1}}^{M_{\rm h2}}\left<M_{\rm HI}|M_{\rm h}\right> n(M_{\rm h})d M_{\rm h}}{\int_{M_{\rm h1}}^{M_{\rm h2}}n(M_{\rm h})dM_{\rm h}}.
    \end{equation}
As will be shown below, the large halo mass bin size will smooth the peak feature of the \hi-halo mass relation at the low mass end, which is more apparent in halos of higher richness (see Figure~2 of \citetalias{Guo2020}). As the observed \hi-halo mass relation is only limited to halos above $10^{11}\msunh$, the low-mass behavior of $\langle M_{\rm HI}|M_{\rm h}\rangle$ can be well constrained with $\Omega_{\rm HI}$. The contribution to \hi\ mass from halos below $10^{11}\msunh$ is estimated to be around 30\% in \citetalias{Guo2020}, which cannot be simply ignored.   

\subsubsection{H{\scriptsize I} Clustering}\label{subsec:hiclustering}
The spatial clustering of \hi\ gas can provide additional constraints on the halo assembly bias effect, as shown in \citetalias{Guo2017} using the 70\% complete sample of ALFALFA. However, we only have the total \hi\ mass information from the \hi-halo mass relation, without the accurate distribution of \hi\ mass within the halos or the measurements of \hi\ gas in subhalos of different masses. As demonstrated in \cite{Guo2021}, the galaxy \hi\ mass generally depends on the stellar mass and star formation rate, so baryon physics needs to be carefully taken into account. It is thus difficult to accurately model the clustering measurements of galaxies with different \hi\ mass thresholds as in \citetalias{Guo2017}.    

To the first order, it is easier to measure and model the clustering of the \hi\ gas itself, by assigning the \hi\ mass as the weight for each observed galaxy in the ALFALFA survey, analogous to the measurement of dark matter clustering. Since the ALFALFA galaxy sample selection depends on both the \hi\ flux and line width, special care must be taken to ensure that accurate clustering measurements are made. We follow the correction method of \citetalias{Guo2017} by assigning each galaxy pair with an additional weight of the effective volume ($V_{\rm eff}$) probed by the two galaxies \citep[see Appendix~B of][for more details]{Martin2010}.  

We use the Landy-Szalay estimator \citep{Landy1993} to measure the redshift-space 3D two-point correlation function $\xi(r_{\rm p},r_{\rm\pi})$, where $r_{\rm\pi}$ and $r_{\rm p}$ are the separations of galaxy pairs along and perpendicular to the line-of-sight, respectively, $\xi(r_{\rm p},r_{\rm\pi})={\rm (DD-2DR+RR)/RR}$. The galaxy pair counts of data-data ($\rm{DD}$), data-random ($\rm{DR}$), and random-random ($\rm{RR}$) are calculated as follows,
\begin{eqnarray}
\rm{DD}(r_{\rm p},r_{\rm\pi})&=&\sum_{(i,j)\in V_{ij}}\frac{M_{{\rm HI},i}M_{{\rm HI},j}}{V_{ij}} \label{eq:dd}, \\
\rm{DR}(r_{\rm p},r_{\rm\pi})&=&\sum_{(i,j)\in V_{ij}}\frac{M_{{\rm HI},i}M_{{\rm HI},j}}{V_{ij}} \label{eq:dr},\\
\rm{RR}(r_{\rm p},r_{\rm\pi})&=&\sum_{(i,j)\in V_{ij}}\frac{M_{{\rm HI},i}M_{{\rm HI},j}}{V_{ij}}, \label{eq:rr}
\end{eqnarray}
where $V_{ij}={\rm min}(V_{{\rm eff},i},V_{{\rm eff},j})$, with $V_{{\rm eff},i}$ and $V_{{\rm eff},j}$ being the effective volumes accessible to the $i^{\rm th}$ and $j^{\rm th}$ galaxies, respectively. Similarly, we also compute the redshift-space two-point correlation function $\xi(s)$, with $s$ being the redshift-space pair separation.

To reduce the effect of redshift-space distortion (RSD), we also measure the projected two-point correlation function $w_{\rm p}(r_{\rm p})$, 
\begin{equation}
w_{\rm p}(r_{\rm p})=\int_{-r_{\pi,{\rm max}}}^{r_{\pi,{\rm max}}} \xi(r_{\rm p},r_{\rm\pi})dr_{\rm\pi}. \label{eq:wp}
\end{equation}
where $r_{\pi,{\rm max}}$ is set to be $20\mpchi$ as in \citetalias{Guo2017}. The residual RSD effect can be further accounted for in our model. 

Our galaxy sample comes from the 100\% complete catalog of the ALFALFA survey \citep{Haynes2018}. The final sample includes 22330 galaxies with reliable \hi\ mass measurements, covering 6518~$\deg^2$ in the redshift range of $0.0025<z<0.06$. The random catalog is constructed similarly as in \citetalias{Guo2017}, with the weights assigned by randomly selecting from the galaxy sample. We will use the measurements of $w_{\rm p}(r_{\rm p})$ and $\xi(s)$ to constrain the \hi-halo mass relation, with logarithmic $r_{\rm p}$ bins of a constant width $\Delta\log r_{\rm p}=0.2$ ranging from $0.13$ to $12.92\mpchi$ (the same for $s$ bins) and linear $r_{\rm\pi}$ bins of width $\Delta r_{\rm\pi}=2\mpchi$ from 0 to 20$\mpchi$. The error covariance matrices are estimated using the jackknife re-sampling method of 100 subsamples. However, as we lack the information of \hi\ mass distribution within the halo, we only use the \hi\ clustering measurements above $1\mpchi$ to avoid the scales dominated by the one-halo contribution. The measurements of $w_{\rm p}(r_{\rm p})$ and $\xi(s)$ are shown as the open circles in the middle and right panels of Figure~\ref{fig:obs}, respectively.

In summary, for the observational data, we have one data point in $\Omega_{\rm HI}$, 10 data points in $\langle M_{\rm HI}|M_{\rm h}\rangle$, 6 data points in $w_{\rm p}(r_{\rm p})$ and 6 data points in $\xi(s)$. The total number of data points is 23. 
    
\subsection{Theoretical Models}\label{sec:model}
\subsubsection{H{\scriptsize I}-halo Relation}
The observed $\langle M_{\rm HI}|M_{\rm h}\rangle$ as shown in Figure~\ref{fig:obs} is not just a smoothly increasing function of $M_{\rm h}$. There is an apparent bump feature around $M_{\rm h}\sim10^{11.5}\msunh$, which is not caused by the measurement errors of the \hi\ stacking, since such a feature becomes more significant for halos of higher richness. As discussed in \citetalias{Guo2020}, it is possibly related to the virial shock-heating and active galactic neuclei (AGN) feedback happening in halos around this mass. Therefore, we propose a flexible, two-component model for $\langle M_{\rm HI}|M_{\rm h}\rangle$, as a combination of a lognormal distribution ($f_{\rm gau}$) at the low mass end and a double power-law (hereafter DPL) distribution $f_{\rm dpl}$, 
\begin{eqnarray}
        &&f(M_{\rm HI}|M_{\rm h})= f_{\rm gau}(M_{\rm h}) + f_{\rm dpl}(M_{\rm h}) \label{eq:f_Mh}\\
        &&f_{\rm gau}/(\msunh) = 10^{A_{\rm gau}\exp{[-(\log (M_{\rm h}/M_{\rm gau})/\sigma_{\rm gau})^2}]}\label{eq:gauss}\\
        &&f_{\rm dpl} = A_{\rm dpl}M_{\rm h}/[(M_{\rm h}/M_{\rm dpl})^{-\alpha}+(M_{\rm h}/M_{\rm dpl})^\beta]\label{eq:dpl}        
\end{eqnarray}
where $A_{\rm gau}$, $M_{\rm gau}$, $\sigma_{\rm gau}$, $A_{\rm dpl}$, $M_{\rm dpl}$, $\alpha$ and $\beta$ are the model parameters. As will be shown below, such a model is flexible enough to fit the observational measurements, as well as the simulation predictions. 

Apart from the halo mass dependence, as we mentioned above, the additional dependence of $M_{\rm HI}$ on the halo assembly history is required to explain the observed clustering measurements. Following \citetalias{Guo2017}, we investigate three halo properties in this paper --- formation time ($a_{1/2}$), concentration parameter ($c_{\rm vir}$) and spin parameter ($\lambda$), as commonly used. The final theoretical model for the \hi-halo relation is formulated as
\begin{equation}
    \langle M_{\rm HI}|M_{\rm h}, P\rangle=f(M_{\rm HI}|M_{\rm h})\cdot P^\gamma,\label{eq:mhi}
\end{equation}
where $\gamma$ is the model parameter and the parameter $P$ can be either $a_{1/2}$, $c_{\rm vir}$ or $\lambda$. The simple power-law dependence on $P$ is motivated by the results in the hydrodynamic simulations, and the fitting to observation seems quite reasonable.

\subsubsection{H{\scriptsize I} Clustering Model}
The model prediction for the cosmic \hi\ abundance $\Omega_{\rm HI}$ can be directly obtained by integrating Eq.~(\ref{eq:omegahi}), where the input of halo mass function $n(M_{\rm h})$ is necessary. Since we also need to include the secondary halo parameters in Eq.~(\ref{eq:mhi}), the analytical halo model which assumes a specific halo mass function form is not directly applicable. We follow the strategy of \citetalias{Guo2017} by directly populating the dark matter halos in the $N$-body simulations using Eq.~(\ref{eq:mhi}). Then $\Omega_{\rm HI}$ can be obtained by summing up the \hi\ mass in the whole simulation volume. 

We use the dark matter halo catalog from the Small MultiDark simulation of Planck cosmology \citep[SMDPL;][]{Klypin2016}, covering a volume of $400^3h^{-3}{\rm Mpc}^3$ and assuming the cosmological parameters of $\Omega_m=0.307$, $\Omega_b=0.048$, $n_s=0.96$, $h=0.678$ and $\sigma_8=0.823$. The particle mass resolution is $9.6\times10^7\msunh$, which is good enough to fully resolve the host halos of \hi-rich galaxies. The halos are identified by the \texttt{ROCKSTAR} halo finder \citep{Behroozi2013} and we only apply our model to the simulation output at $z=0$.

The halo properties of formation time, concentration and spin parameter are also available from the \texttt{ROCKSTAR} halo catalog. The halo formation time $a_{1/2}$ is defined as the scale factor at which the halo mass first reaches half of the peak value over the whole merger history. The concentration parameter $c_{\rm vir}$ is the ratio between halo virial radius and scale radius. The spin parameter $\lambda$ is calculated using the definition of \cite{Peebles1969}.

Since the ALFALFA survey is limited to the local universe with $z<0.06$, the sample variance effect for the clustering measurements is still severe and will cause a systematic underestimation of the large-scale clustering measurements. Therefore, we correct for this finite volume effect, known as the ``integral constraint'', following the method presented in Section~4.4 of \citetalias{Guo2017}. Briefly speaking, for each run of the model parameters in Eq.~(\ref{eq:mhi}), we construct 64 mock galaxy catalogs from the simulation box with the same geometry as the ALFALFA survey, calculate the clustering measurements (with the RSD effect automatically included using halo peculiar velocities), and use the average of the 64 mocks as the model prediction.  

\subsubsection{Model Fitting}
To find the bestfit model parameters to the observational data described in Section~\ref{sec:data}, we apply the Monte Carlo Markov Chain (MCMC) technique, using the Bayesian inference tool of \texttt{MultiNest} \citep{Feroz2009}. The likelihood surface is determined by $\chi^2$, 
\begin{eqnarray}
    &&\chi^2=\chi^2_1+\chi^2_2+\chi^2_3\\        
    &&\chi^2_1=(\Omega_{\rm HI}-\Omega^*_{\rm HI})^2/\sigma_{\Omega_{\rm HI}}^2\\
    &&\chi^2_2=(\langle M_{\rm HI}|M_{\rm h}\rangle-\langle M_{\rm HI}|M_{\rm h}\rangle^*)^2/\sigma_{\langle M_{\rm HI}|M_{\rm h}\rangle}^2\\
    &&\chi^2_3=(\mathbf{\xi_{\rm all}}-\mathbf{\xi^*_{\rm all}})^T\mathbf{C}^{-1}(\mathbf{\xi_{\rm all}}-\mathbf{\xi^*_{\rm all}})
\end{eqnarray}
where the data vector $\mathbf{\xi_{\rm all}}=[w_{\rm p}(r_{\rm p}),\xi(s)]$ (i.e., the combination of $w_{\rm p}(r_{\rm p})$ and $\xi(s)$), and $\mathbf{C}$ is the full error covariance matrix for $\xi_{\rm all}$. The quantity with (without) a superscript `*' is the one from the measurement (model). The cross-correlation between $w_{\rm p}(r_{\rm p})$ and $\xi(s)$ has been fully accounted for in $\mathbf{C}$. We also assume that the measurements of $\Omega_{\rm HI}$, $\langle M_{\rm HI}|M_{\rm h}\rangle$, and $\xi_{\rm all}$ are independent of each other. Also, note that the error $\sigma_{\langle M_{\rm HI}|M_{\rm h}\rangle}$ was estimated for the mean of \hi\ mass for a given halo mass bin using the jackknife method, and therefore smaller than the error that would be estimated as the standard deviation of the sample of \hi\ mass at individual halos. 

With 23 data points and 8 model parameters, the final degree-of-freedom (dof) for the model fitting is 15, unless otherwise noted.
    
\section{Results}
\label{sec:observational result}
\subsection{Bestfit Models}
    \begin{figure*}
        \centering
        \includegraphics[width = \textwidth]{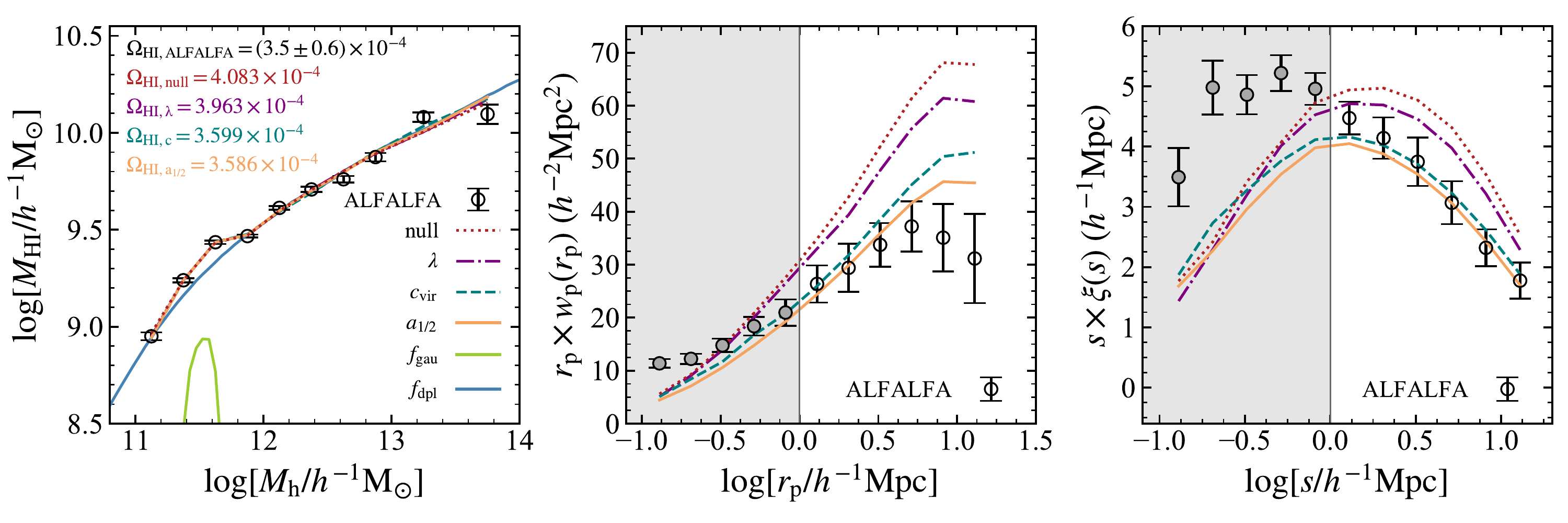}
        \caption{Best-fitting models to the observational results of ALFALFA including \hi-halo mass average relation $\langle M_{\rm HI}|M_{\rm h}\rangle$ (left panel), projected correlation function $w_{\rm p}(r_{\rm p})$ (middle panel) and correlation function $\xi(s)$ (right panel). Black circles represent the observational data used in our fitting with error bars estimated using the jackknife method, while colored lines denote the predictions using Eq.(\ref{eq:mhi}), including halo spin parameter (``$\lambda$''), halo concentration (``$c_{\rm vir}$''), and halo formation time (``$a_{1/2}$''). In particular, the red dotted line denotes the fitting results without secondary dependence of halo parameters according to Eqs.(\ref{eq:f_Mh}-\ref{eq:dpl}) (``null''). In the left panel, the light green is the analytical Gaussian-like function (``$f_{\rm gau}$''), and the dark blue is the double power law (``$f_{\rm dpl}$"). Data points in the grey shaded regions in the middle and right panels are not used in fitting, because these regions signify the fine structures of galaxies where \hi\ gas can not be assigned simply as a function of halo mass.}
        \label{fig:obs}
    \end{figure*}
    
\begin{table*}
    \begin{threeparttable}
    \caption{Bestfit Model Parameters\tnote{a} for Observational Measurements}
    \label{tab:coeff}
    \centering
    \begin{tabular}{ccccccccccc}
        \toprule
        P\tnote{b} & $\chi^2/{\rm dof}$\tnote{c} & $A_{\rm gau}$ & $\log M_{\rm gau}$ & $\sigma_{\rm gau}$ & $\log A_{\rm dpl}$ & $\log M_{\rm dpl}$ & $\alpha$ & $\beta$ & $\gamma$ \\
        \midrule
        null & $52.74/16$ & $9.17^{+ 0.03}_{-0.27}$ & $11.52^{+0.03}_{-0.00}$ & $0.38^{+0.38}_{-0.01}$ & $-1.93^{+0.06}_{-0.05}$ & $11.45^{+0.11}_{-0.21}$ & $0.09^{+0.17}_{-0.06}$ & $0.71^{+0.02}_{-0.05}$ & 0 \\
        $\lambda$ & $42.41/15$ & $5.51^{+1.23}_{-0.68}$ & $11.51^{+0.02}_{-0.01}$ & $0.34^{+0.25}_{-0.07}$ & $-5.64^{+1.34}_{-0.62}$ & $11.66^{+0.08}_{-0.24}$ & $0.01^{+0.16}_{-0.06}$ & $0.81^{+0.02}_{-0.07}$ & $-2.21^{+0.73}_{-0.34}$ \\
        $c_{\rm vir}$ & $29.08/15$ & $11.98^{-0.05}_{-0.84}$ & $11.51^{+0.05}_{-0.03}$ & $0.51^{+0.38}_{-0.08}$ & $1.17^{+0.12}_{-0.79}$ & $11.18^{+0.11}_{-0.08}$ & $0.41^{+0.39}_{-0.26}$ & $1.00^{+0.02}_{-0.09}$ & $-3.12^{+0.86}_{-0.12}$\\
        $a_{1/2}$ & $27.23/15$ & $10.60^{+0.30}_{-0.36}$ & $11.51^{+0.03}_{-0.01}$ & $0.45^{+0.32}_{-0.09}$ & $-0.34^{+0.42}_{-0.24}$ & $11.23^{+0.05}_{-0.15}$ & $0.24^{+0.50}_{-0.15}$ & $0.88^{+0.03}_{-0.05}$ & $5.60^{+2.26}_{-1.12}$\\
        \bottomrule
    \end{tabular}
    	\medskip
	{\bf Note.} 
	\begin{tablenotes}
	    \item [a] The parameters $M_{\rm gau}$ and $M_{\rm dpl}$ are in units of $\msunh$. Other parameters are unitless. 
	    \item [b] The first model (``null'') assumes no secondary dependence of halo parameters, i.e.\ $\gamma$ is set to zero. Other models include secondary dependence of \hi\ mass on halo spin parameter (``$\lambda$''), halo concentration (``$c_{\rm vir}$''), and halo formation time (``$a_{1/2}$''), respectively. The $a_{1/2}$-model shows the best agreement with the observation data. 
	    \item [c] Note that the error for ${\langle M_{\rm HI}|M_{\rm h}\rangle}$ was estimated for the mean of \hi\ mass for a given halo mass bin using the jackknife method. Therefore $\chi^2$ here is overestimated compared to the value it would be if the error would be estimated as the standard deviation of the sample of \hi\ mass at individual halos. 
	\end{tablenotes}
	\end{threeparttable}
\end{table*}

To test the effect of halo assembly bias, we first fit the observational measurements without inclusion of any dependence on the secondary halo parameters (i.e.\ $\gamma=0$ in Eq.~\ref{eq:mhi}). The bestfit model is shown as the red dotted lines in Figure~\ref{fig:obs} (indicated as `null'), with $\chi^2/{\rm dof}$ of $52.74/16$. (Note that in this case there are 7 free model parameters and therefore the dof is 16).  Although it fits the relation of $\langle M_{\rm HI}|M_{\rm h}\rangle$ well with the adopted functional form of Eq.~(\ref{eq:f_Mh}), the clustering measurements of $w_{\rm p}(r_{\rm p})$ and $\xi(s)$ are significantly overestimated. The predicted value of $\Omega_{\rm HI} = 4.08\times10^{-4}$ is also overestimated by 1$\sigma$, as indicated in the top of the left panel. Assuming a $\chi^2$ distribution, this model can be strongly rejected, as the best-fitting $\chi^2$ is 6.49$\sigma$ away from the mean value of 16. It thus strongly supports the necessity of including the halo assembly bias effect in the modeling of the \hi\ content in halos, confirming the results of \citetalias{Guo2017} and \citetalias{Guo2020}. 

The best-fitting models of including $\lambda$, $c_{\rm vir}$ and $a_{1/2}$ are shown as the dot-dashed, dashed, and solid lines in Figure~\ref{fig:obs}, respectively. They are indistinguishable from each other in the fit to $\langle M_{\rm HI}|M_{\rm h}\rangle$, as it does not provide any constraints on the parameter $\gamma$ in Eq.~(\ref{eq:mhi}). However, large differences in the models are seen in the clustering measurements of $w_{\rm p}(r_{\rm p})$ and $\xi(s)$. The $\lambda$-model apparently over-predicts the large-scale clustering measurements and the best-fitting $\chi^2$ is about 5$\sigma$ from the mean value of 15. The $a_{1/2}$ and $c_{\rm vir}$ models agree with each other, with the $a_{1/2}$ model showing slightly better agreement with the observation. These results are fully consistent with those of \citetalias{Guo2017}. In \citetalias{Guo2017}, using $\lambda$ as the secondary halo parameter in an improved abundance matching model overestimates the large-scale clustering measurements. The empirical relation for $\langle M_{\rm HI}|M_{\rm h}\rangle$ proposed in this paper provides more quantitative description between the \hi\ mass and halo mass. 

For the two components ($f_{\rm gau}$ and $f_{\rm dpl}$ in Eq.~\ref{eq:f_Mh}) in the proposed $\langle M_{\rm HI}|M_{\rm h}\rangle$ function, their individual contributions to the best-fitting $a_{1/2}$-model are shown as the green and blue solid lines in the left panel of Figure~\ref{fig:obs}, respectively. Most \hi\ mass is contributed by the $f_{\rm dpl}$ component, and $f_{\rm gau}$ acts as a sharp peak around the halo mass of $10^{11.5}\msunh$. We note that Eq.~(\ref{eq:f_Mh}) is only an empirical formula for the average \hi\ mass in the halo. However, the deviation from the smooth increasing of $M_{\rm HI}$ with $M_{\rm h}$ at the scale of $M_{\rm gau}$ encodes the information about the physical processes involved, including virial shock-heating and AGN feedback for halos more massive than $M_{\rm gau}$ (see \citetalias{Guo2020}). 

While we focus on the total \hi\ mass in the halo, it is possible to separate the contributions from central and satellite galaxies. As shown in \citetalias{Guo2020}, the total \hi\ mass is dominated by the central galaxies with $M_{\rm h}<10^{13}\msunh$. This is different from our division of $f_{\rm gau}$ and $f_{\rm dpl}$, since $f_{\rm dpl}$ dominates at all mass scales. As we noted above, the main difficulty to model the  contributions from central and satellite galaxies is that the galaxy \hi\ mass significantly depends on the details in baryonic processes, such as star-formation and feedback \citep{Guo2021}. We will defer a physical model for the total \hi\ mass in the halo and its evolution to future work. 

\begin{figure}
    \centering
    \includegraphics[width = 0.45\textwidth]{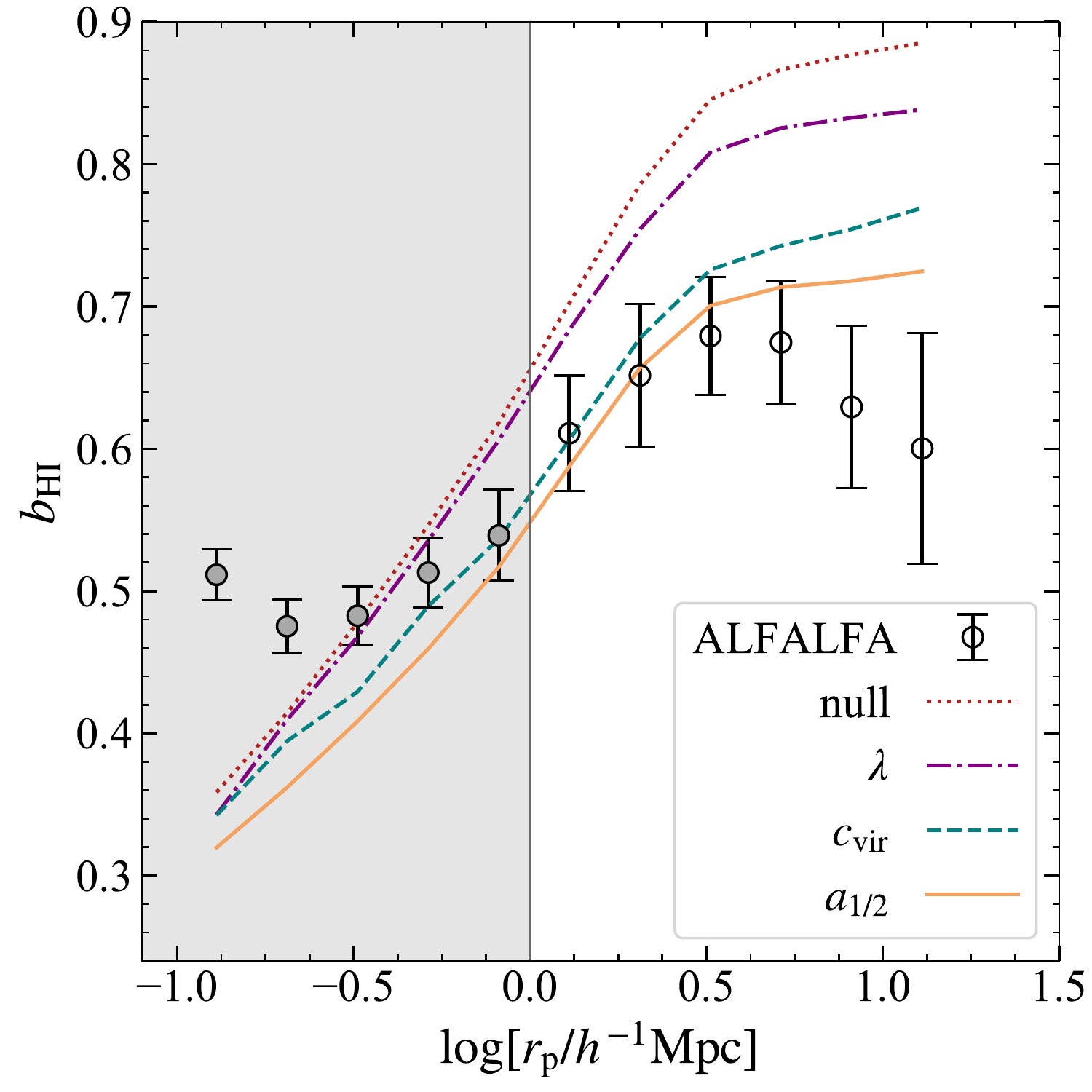}
    \caption{The \hi\ bias $b_{\rm HI}\left(r\right)$ predicted by the bestfit models with the ALFALFA data (with the same color types and line types as in Figure~\ref{fig:obs}), in comparison with the observational results (black circles) with error bars estimated using the jackknife method. Data points at $r_{\rm p} < 1 \mpchi$, as shown in the grey shaded region, are not used, due to small scale effects.} 
    \label{fig:bias_obs}
\end{figure}

We can quantify the scale-dependence of the clustering measurements using the \hi\ bias, $b_{\rm HI}(r_{\rm p})$, defined as, 
\begin{equation}
    b_{\rm HI}(r_{\rm p})=\sqrt{\frac{w_{\rm p}(r_{\rm p})}{w_{\rm p,dm}(r_{\rm p})}},
\end{equation}
where $w_{\rm p,dm}(r_{\rm p})$ is the projected two-point correlation function for the dark matter. It is measured in the same way as $w_{\rm p}$ by including the effect of integral constraint. The resulting $b_{\rm HI}(r_{\rm p})$ measurements are shown in Figure~\ref{fig:bias_obs}. The best-fitting $a_{1/2}$-model has a large-scale \hi\ bias of $b_{\rm HI}\sim 0.7$, while the ``null''-model has a significantly larger bias of $b_{\rm HI}\left(r\right) \sim 0.87$. The observed value of \hi\ bias is much smaller than the ``null''-model prediction, which indicates that the neutral hydrogen at $z\simeq 0$ is distributed in the low-mass halos as well as late-forming (or low-concentration) halos. 

Without including the secondary halo parameters, \cite{Obuljen2019} obtained a constraint of the \hi\ bias,  $b_{\rm HI}=0.878^{+0.022}_{-0.023}$, by fitting to the \hi\ clustering measurements and $\Omega_{\rm HI}$ of the ALFALFA galaxy sample, which is consistent with our null model here. However, their measured large-scale \hi\ bias is larger than ours due to the sample selection effect of ALFALFA survey, where the observed sample is dominated by the galaxies of $M_{\rm HI}\sim10^{9.8}$ with a bias of $b\sim0.86$ (see Figure~8 of \citetalias{Guo2017}). When the target selection is suitably taken into account with the effective volume weight as in Section~\ref{subsec:hiclustering}, the \hi\ bias is reduced due to the greater contribution from galaxies with lower $M_{\rm HI}$.  

\begin{figure*}
    \centering
    \includegraphics[width = \textwidth]{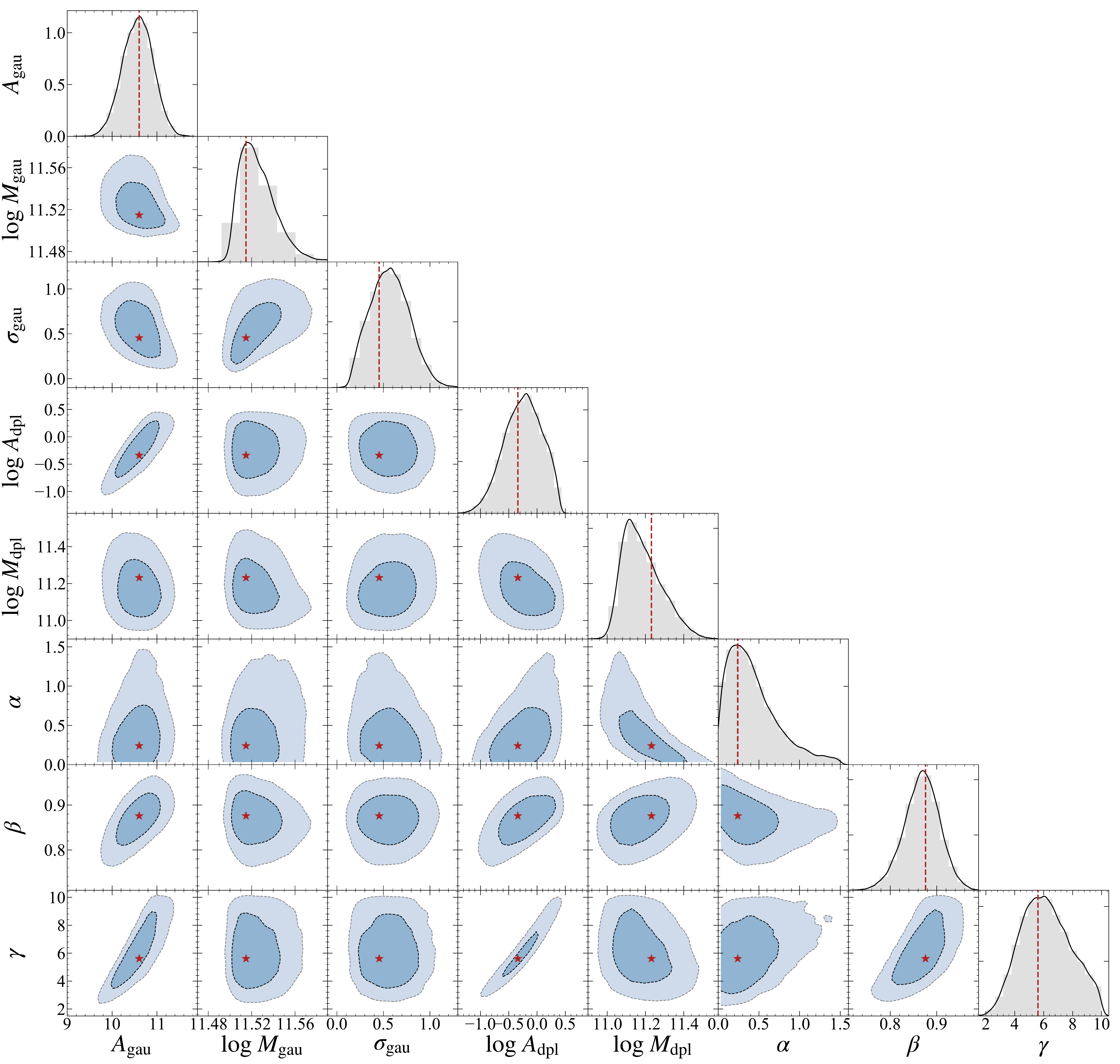}
    \caption{Posterior parameter distributions for the bestfit $a_{1/2}$-model that agrees best with the observations. We show the probability distribution function (PDF) of each parameter in grey histograms, and the correlation between different parameters with 68\% (95\%) confidence level in blue (light-blue) contours. The bestfit models are marked as red stars in contours and vertical red dashed lines in histograms. Note that ``parameters'' here refer to the coefficients in Eqs.~(\ref{eq:f_Mh})---(\ref{eq:mhi}), and should not be confused with the aforementioned ``secondary parameters''.}
    \label{fig:param_dist}
\end{figure*}

We show the bestfit parameters for different models in Table~\ref{tab:coeff}. Due to the observed low \hi\ bias, the best-fitting model prefers to populate halos with lower bias values. As shown in Figure~4 of \cite{Sato2019}, for halos of $M_{\rm h}<10^{13}\msunh$, the late-forming and low-$c_{\rm vir}$ halos would have lower bias relative to their early-forming and high-$c_{\rm vir}$ counterparts. Therefore, the bestfit value of $\gamma$ in Table~\ref{tab:coeff} is positive for the $a_{1/2}$-model and negative for the $c_{\rm vir}$-model. 

The situation for the $\lambda$-model is slightly more complicated, since the relative bias between the high-$\lambda$ and low-$\lambda$ halos would switch at around $10^{11.5}\msunh$. However, our model only assumes a constant value of $\gamma$ for the whole halo mass range. The negative value of bestfit $\gamma$ for the $\lambda$-model indicates that the neutral hydrogen prefers to populate the low-spin halos, which have lower bias for $M_{\rm h}>10^{11.5}\msunh$. However, the relative difference of $b_{\rm HI}$ between the null model and the $a_{1/2}$-model value is as large as $0.87/0.7 -1 \approx 25\%$. For halos with $M_{\rm h}<10^{12}\msunh$ (which compass most of the cosmic \hi), the relative bias between the low-$\lambda$ (or high-$\lambda$) halos and the ensemble average at the same mass is less than 10\% \citep{Sato2019}. Thus, even if a $M_{\rm h}$-dependent $\gamma$ parameter would be introduced to the $\lambda$-model, $b_{\rm HI}$ would not be reduced by 25\% in the null model. As seen in Figure~\ref{fig:bias_obs}, the $\lambda$-model only has a slightly decreased large-scale bias compared to the null model. We will further discuss the effect of halo spin in the next section.   
 
We show the posterior parameter distributions for the $a_{1/2}$-model in Figure~\ref{fig:param_dist}. Most correlations between different parameters are weak, which implies that these parameters are mostly independent of each other. However, there are strong correlations between $\gamma$ and $A_{\rm gau}$, and between $\gamma$ and $A_{\rm dpl}$, as expected due to Eq.~(\ref{eq:mhi}). 

Since $\gamma$ can be constrained from the clustering measurements, these correlations imply that the \hi-halo mass distribution $f(M_{\rm HI}|M_{\rm h})$ is not independent of the halo formation time distribution. This is also confirmed by the fact that for all coefficients (except for $M_{\rm gau}$) in Table~\ref{tab:coeff}, their values vary significantly among different models. It means that even if the observed $\langle M_{\rm HI}|M_{\rm h}\rangle$ is well fit by different models, their \hi\ mass distributions in each halo mass bin vary from each other. Such differences are reflected in halos below the lower limit $10^{11}\msunh$ of the observed \hi-halo mass relation, causing the differences in the predictions of $\Omega_{\rm HI}$.

\subsection{\rm{H{\scriptsize I}} Mass Distributions}

\begin{figure}
	\centering
	\includegraphics[width=0.45\textwidth]{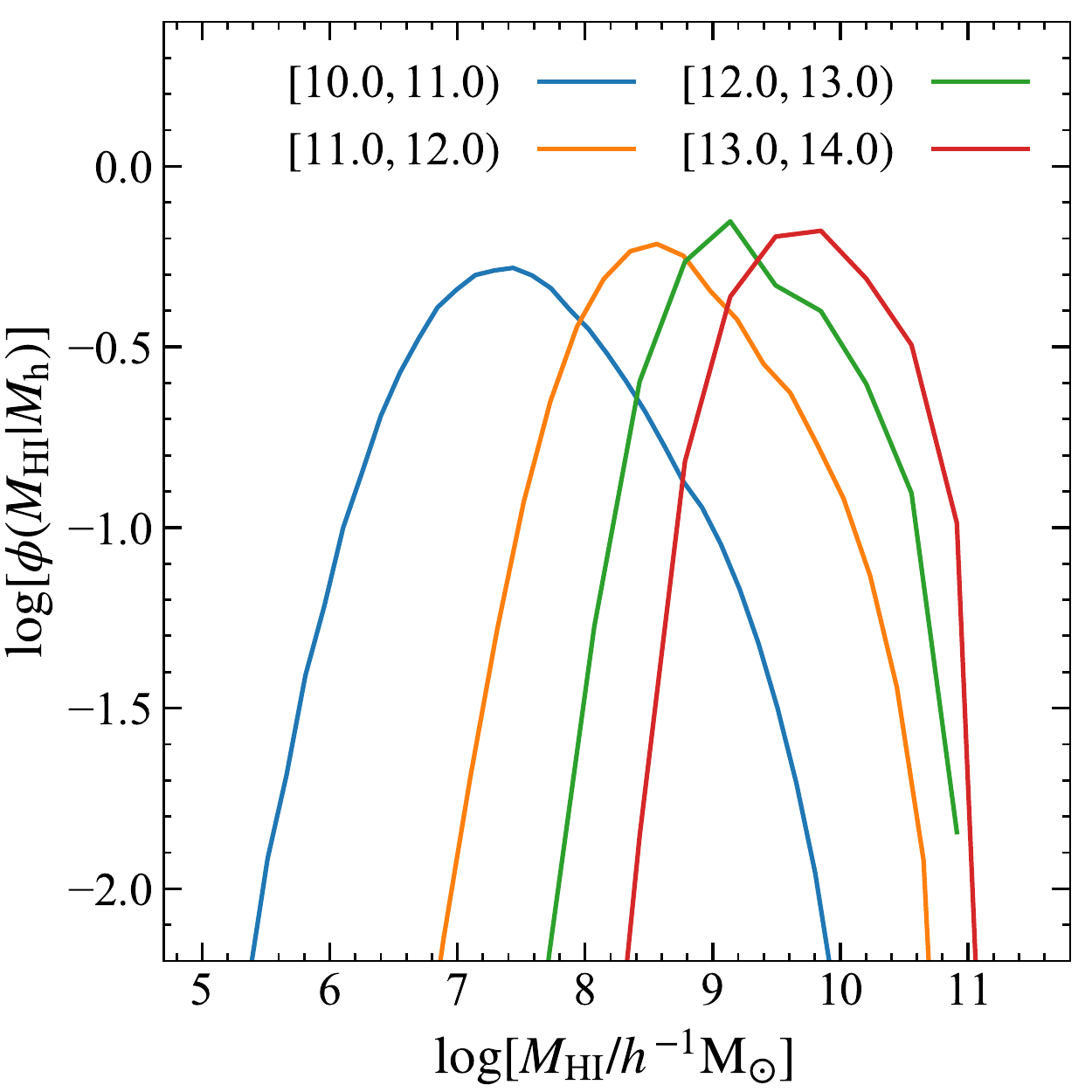}
	\caption{Probability distribution function of \hi\ mass, $\phi(M_{\rm HI}|M_{\rm h})$, for the bestfit $a_{\rm 1/2}$-model for the SMDPL N-body simulation, in different halo mass bins, i.e. at  $10.0\le \log(M_{\rm h}/\msunh) < 11.0$ (blue), $11.0\le \log(M_{\rm h}/\msunh) < 12.0$ (orange), $12.0\le \log(M_{\rm h}/\msunh) < 13.0$ (green), and $13.0\le \log(M_{\rm h}/\msunh) < 14.0$ (red), respectively. These distributions look close to the log-normal distribution, with average scatter decreasing from 0.7 to 0.5 when $M_{\rm h}$ increases from $10^{10} \msunh$ to $10^{14} \msunh$.}
	\label{fig:scatter}
\end{figure}

When we fit the \hi-halo mass relation, we only use the information of average \hi\ mass in each $M_{\rm h}$ bin, i.e. $\langle M_{\rm HI}|M_{\rm h}\rangle$, without assuming the probability distribution of \hi\ mass in halos, $\phi(M_{\rm HI}|M_{\rm h})$. However, the secondary parameter dependence in Eq.~(\ref{eq:mhi}) actually encodes the information of $\phi(M_{\rm HI}|M_{\rm h})$. Thus, with the bestfit model parameters in Table~\ref{tab:coeff}, we can construct mock catalogs from the simulations and quantify $\phi(M_{\rm HI}|M_{\rm h})$.

We adopt the best-fitting $a_{1/2}$-model and show in Figure~\ref{fig:scatter} the distributions of $\phi(M_{\rm HI}|M_{\rm h})$ in different halo mass bins. While they look close to the log-normal distributions, the distributions are increasingly skewed for less massive halos. This is likely caused by the inclusion of the formation time dependence in $\phi(M_{\rm HI}|M_{\rm h})$. In principle, $\phi(M_{\rm HI}|M_{\rm h})$ can be expanded as $\phi(M_{\rm HI}|M_{\rm h},a_{1/2})$. However, in our model, $M_{\rm HI}$ is completely determined by $M_{\rm h}$ and $a_{1/2}$, without any scatter. The current observational data of ALFALFA is not large enough to fully constrain $\phi(M_{\rm HI}|M_{\rm h},a_{1/2})$. We find from Figure~\ref{fig:scatter} that the average scatter of $M_{\rm HI}$ in different $M_{\rm h}$ bins decreases from $0.7$ to $0.5$ when $M_{\rm h}$ increases from $10^{10}\msunh$ to $10^{14}\msunh$.

\begin{figure}
	\centering
	\includegraphics[width = 0.45\textwidth]{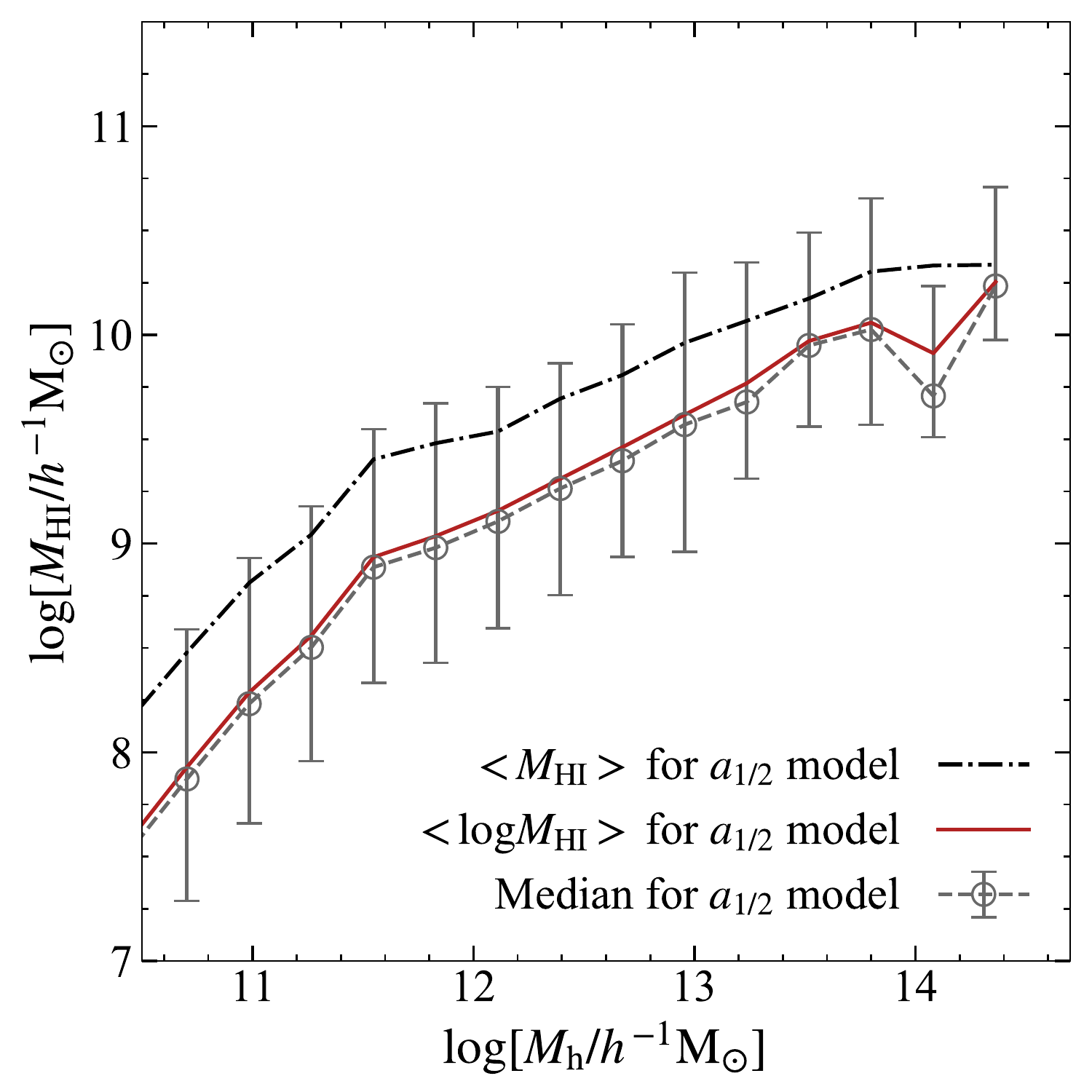}
	\caption{Different approaches to represent the average \hi, including ${\rm log} \left<M_{\rm HI}\right>$ (black dot-dashed line), $\left<{\rm log} M_{\rm HI}\right>$ (red solid line) and median $M_{\rm HI}$ in logarithmic scale (grey dashed line with error bars), all of which are calculated by applying the best-fitting $a_{\rm 1/2}$-model to the TNG100-1-Dark at $z=0$.} 
	\label{fig:log-normal}
\end{figure}

\begin{figure*}
	\centering
	\includegraphics[width = \textwidth]{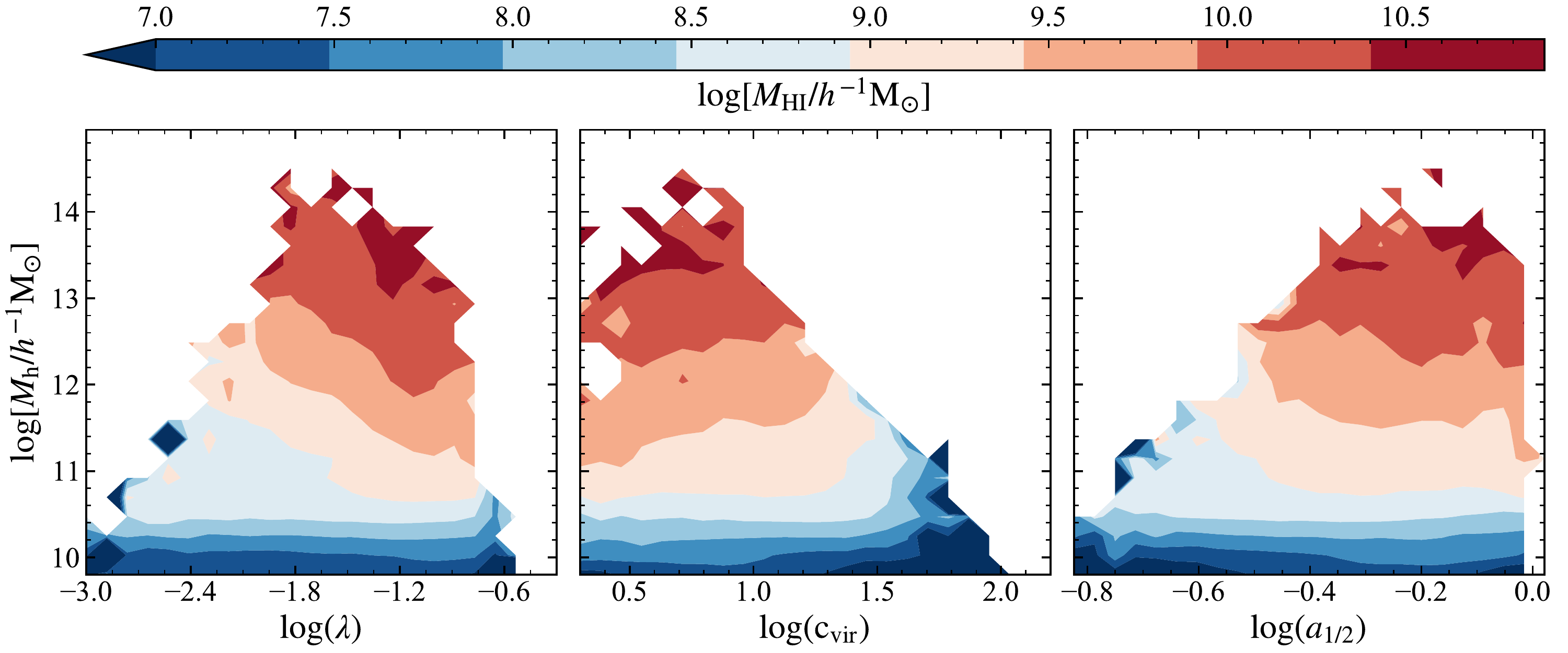}
	\caption{The joint dependence of \hi\ mass on both halo mass and secondary halo parameters in TNG100 simulation at $z=0$. We show the dependence on halo spin parameter $\lambda$ (left), concentration $c_{\rm vir}$ (middle), and formation time $a_{\rm 1/2}$ (right). The \hi\ mass is represented by the color in each panel, with the color bar shown on the top.}
	\label{fig:SD}
\end{figure*}

The measurements of $\phi(M_{\rm HI}|M_{\rm h})$ is sometimes referred to as the conditional \hi\ mass function (CHMF). Nevertheless, our $M_{\rm HI}$ measurements are the total \hi\ mass in halos rather than the \hi\ mass of individual galaxies, as commonly used. Our model predictions can still be compared with the measurements of CHMF built from the galaxies (X.~Li et al. in prep.), providing more constraints to the \hi\ content of galaxies. 

Another application of $\phi(M_{\rm HI}|M_{\rm h})$ is the calculation of \hi\ mass measured in different ways. It is important that the comparison of different studies is made with consistent definitions. There are three commonly used quantities --- $\log\langle M_{\rm HI}\rangle$, $\langle\log M_{\rm HI}\rangle$, and the logarithm of the median values of $M_{\rm HI}$, in different halo mass bins. While the \hi\ stacking measurements of \citetalias{Guo2020} is measured with $\langle M_{\rm HI}\rangle$, it is general to compare the median $M_{\rm HI}$ in theoretical models. They all can be directly calculated from $\phi(M_{\rm HI}|M_{\rm h})$. But for simplicity, we use the same mock catalog as in Figure~\ref{fig:scatter} and the show results of three measurements in Figure~\ref{fig:log-normal}. The errors of median $M_{\rm HI}$ are estimated from the $16^{\rm th}$--$84^{\rm th}$ percentile range. 

We find that the logarithm of the median $M_{\rm HI}$ is very close to $\langle\log M_{\rm HI}\rangle$, but systematically smaller than $\langle M_{\rm HI}\rangle$ by about 0.3--0.5~dex, indicating a log-normal distribution for the \hi\ mass. As expected, the measurement of $\langle M_{\rm HI}\rangle$ would be up-weighted by galaxies with larger $M_{\rm HI}$, causing the difference to the median values. However, their differences are still within the 1$\sigma$ range of the median values, and there is no strong systematic trend of the difference with $M_{\rm h}$, as opposed to some theoretical models \citep[e.g.][]{Calette2021}. We adopt the convention of taking the mean ($\langle M_{\rm HI}\rangle$) throughout this paper.

\section{Comparison with Simulations}
\label{sec:simulation part}

\subsection{The IllustrisTNG Simulation}\label{sec:TNG_Intro}

In attempt to give a theoretical interpretation to our findings, we will investigate in this section the \hi-halo mass relation and its dependence on other halo properties using the IllustrisTNG simulation at $z = 0$.

The IllustrisTNG (hereafter ``TNG'') simulation \citep{Marinacci2018,Naiman2018,Nelson2018,Pillepich2018,Springel2018} is a set of state-of-the-art hydrodynamical simulations, containing a wide range of physical processes of galaxy formation with the same Planck cosmology as SMDPL. In this work, we adopt the simulation set of TNG100-1 with a box size of $110.7$~Mpc and a dark matter particle mass resolution of $7.5\times10^6\msun$. In addition, we also use the dark matter-only simulation of TNG100-1-Dark, to apply our \hi-halo model.

Our modeling of the observational data in the previous section is based on halo catalogs generated with the \texttt{ROCKSTAR} halo finder, which is different from the original halos identified with \texttt{SUBFIND} \citep{Springel_2001} in TNG. For the purpose of fair comparisons, we reprocess the snapshot data of TNG100-1 and TNG100-1-dark to generate halos and subhalos using the \texttt{ROCKSTAR} algorithm. The halo properties of formation time, concentration and spin are also provided by the \texttt{ROCKSTAR} finder. We adopt the post-processed \hi\ mass of TNG subhalos in \cite{Diemer2018}, where five different models for the atomic-to-molecular transition are presented. For simplicity, we employ the ``K13" model for our \hi\ mass estimates. However, the results of other models are similar \citep{Diemer2019} and do not affect the conclusions herein. We note that these \hi\ masses were generated for the \texttt{SUBFIND} subhalos. Thus, we further cross-match the \texttt{SUBFIND} subhalos with our \texttt{ROCKSTAR} subhalos to obtain the final \hi\ mass for each host halo in our catalog.

\subsection{H{\scriptsize I}-halo Relation in TNG}\label{sec:secondary dependence in TNG}

\begin{figure*}
	\centering
	\includegraphics[width = \textwidth]{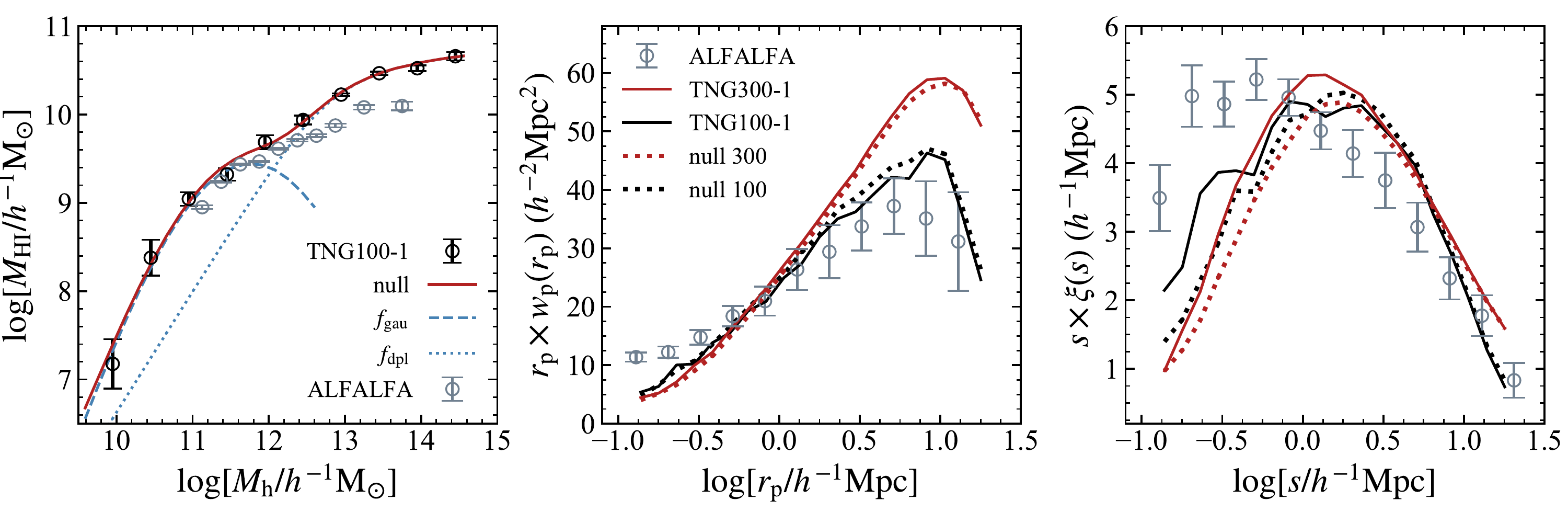}
	\caption{(Left panel) Best-fitting models to the TNG100-1 simulation results at $z=0$ including \hi-halo mass average relation $\langle M_{\rm HI}|M_{\rm h}\rangle$ and $\Omega_{\rm HI}$. Black circles with error bars represent the results of $\langle M_{\rm HI}|M_{\rm h}\rangle$ in the TNG100-1 simulation, and the red solid line denotes the prediction by applying the ``null''-model (i.e.\ without secondary dependence of halo parameters according to Eqs.\ref{eq:f_Mh}-\ref{eq:dpl}) to the TNG100-1-Dark simulation. The blue dashed line is the analytical Gaussian-like function (``$f_{\rm gau}$''), and the blue dotted line is the double power law (``$f_{\rm dpl}$"). For comparison, we show the measurements of ALFALFA with error bars (grey circles). 
	(Middle and right panels) We show the projected correlation function $w_{\rm p}(r_{\rm p})$ (middle) and correlation function $\xi(s)$ (right). Here, we show the results from the TNG300-1 and TNG100-1 simulations (red solid and black solid lines), and the predictions by applying the ``null''-model to the halo data of the TNG300-1 simulation (red dotted line) and to the TNG100-1-Dark simulation (black dotted line), respectively.}
	\label{fig:result}
\end{figure*} 

In Figure~\ref{fig:SD}, we show the joint dependence of $M_{\rm HI}$ on $M_{\rm h}$ and secondary halo parameters in TNG, including halo spin parameter $\lambda$, concentration $c_{\rm vir}$, and formation time $a_{1/2}$. For low-mass halos of $M_{\rm h}<10^{10.8}\msunh$, the dependence of $M_{\rm HI}$ on the secondary parameters is generally very weak, as indicated by the flat contours in Figure~\ref{fig:SD}. This means that for such low-mass halos, the \hi\ mass is mainly determined by the host halo mass. For more massive halos, large \hi\ content tends to be found in high-spin, low-concentration, and young halos. The \hi\ mass shows the strong dependence on $\lambda$ and weak dependence on $c_{\rm vir}$ and $a_{1/2}$. 

Qualitatively, the TNG result that the \hi\ mass depends on both halo mass and the secondary parameters of halo formation history is consistent with our previous conclusion in the observational analysis. Nevertheless, the observations reject the model in which the \hi\ mass depends on halo spin parameter $\lambda$, but show the strong dependence of $M_{\rm HI}$ on $a_{1/2}$ and $c_{\rm vir}$, as shown in Figure~\ref{fig:obs}. That is in conflict with the trend found in TNG. 

In the left panel of Figure~\ref{fig:result}, we show the \hi-halo mass relation in the TNG100-1 simulation. We also display the measurements of the ALFALFA sample as the gray circles. The overall trend of increasing $M_{\rm HI}$ with $M_{\rm h}$ in the TNG100-1 simulations is found in a similar manner to that in the ALFALFA measurements. The low-mass halos of $M_{\rm h}<10^{12}\msunh$ in the TNG simulation have similar content of \hi\ gas to those in the ALFALFA observation. However, the total \hi\ content in more massive halos is apparently over-predicted in the TNG simulation. As discussed in \citetalias{Guo2020}, this discrepancy results from the low-level AGN feedback in TNG, making over-abundant cold gas in the central galaxies. 

We also apply the ``null''-model to the TNG100-1-Dark simulation, and obtain the bestfit model parameters by fitting its prediction of \hi\ abundance $\Omega_{\rm HI}$ and average $\langle M_{\rm HI}|M_{\rm h}\rangle$ relation to the results of the TNG100-1 simulation. The left panel of Figure~\ref{fig:result} shows that the average $\langle M_{\rm HI}|M_{\rm h}\rangle$ relation can be well described by the null-model without secondary halo parameters, as found in the case of fitting to the observational data (see Figure~\ref{fig:obs}). 

In the middle and right panels of Figure~\ref{fig:result}, we show the \hi\ clustering measurements, i.e.\ $w_{\rm p}(r_{\rm p})$ and $\xi(s)$, in the TNG simulations. Note that while the clustering measurements from the TNG100-1 simulation agree with the ALFALFA observations reasonably well, a fair comparison should be made between the ALFALFA (with the sample volume of $3.7\times10^6\,h^{-3}{\rm Mpc}^3$) and the TNG300-1 simulation (with a volume of $8.6\times 10^6\,h^{-3}{\rm Mpc}^3$, or 2.3 times larger than the ALFALFA survey volume), as opposed to that between the ALFALFA and the TNG100-1 simulation (with a volume of $0.42\times10^6\,h^{-3}{\rm Mpc}^3$, or one-ninth of the ALFALFA survey volume), because of the integral constraint effect due to small volume size of a survey or simulation. 
The initial conditions and post-processing of \hi\ and H$_2$ models are exactly the same in the TNG100-1 and TNG300-1 simulations, but Figure~\ref{fig:result} shows that the \hi\ clustering amplitude is suppressed at large scales in the TNG100-1 simulation, compared to the TNG300-1 simulation, which results from the integral constraint effect. The small-scale (below $1\mpchi$) difference between both simulations is likely caused by the resolution effect.

\begin{figure*}
	\centering
	\includegraphics[width = 0.85\textwidth]{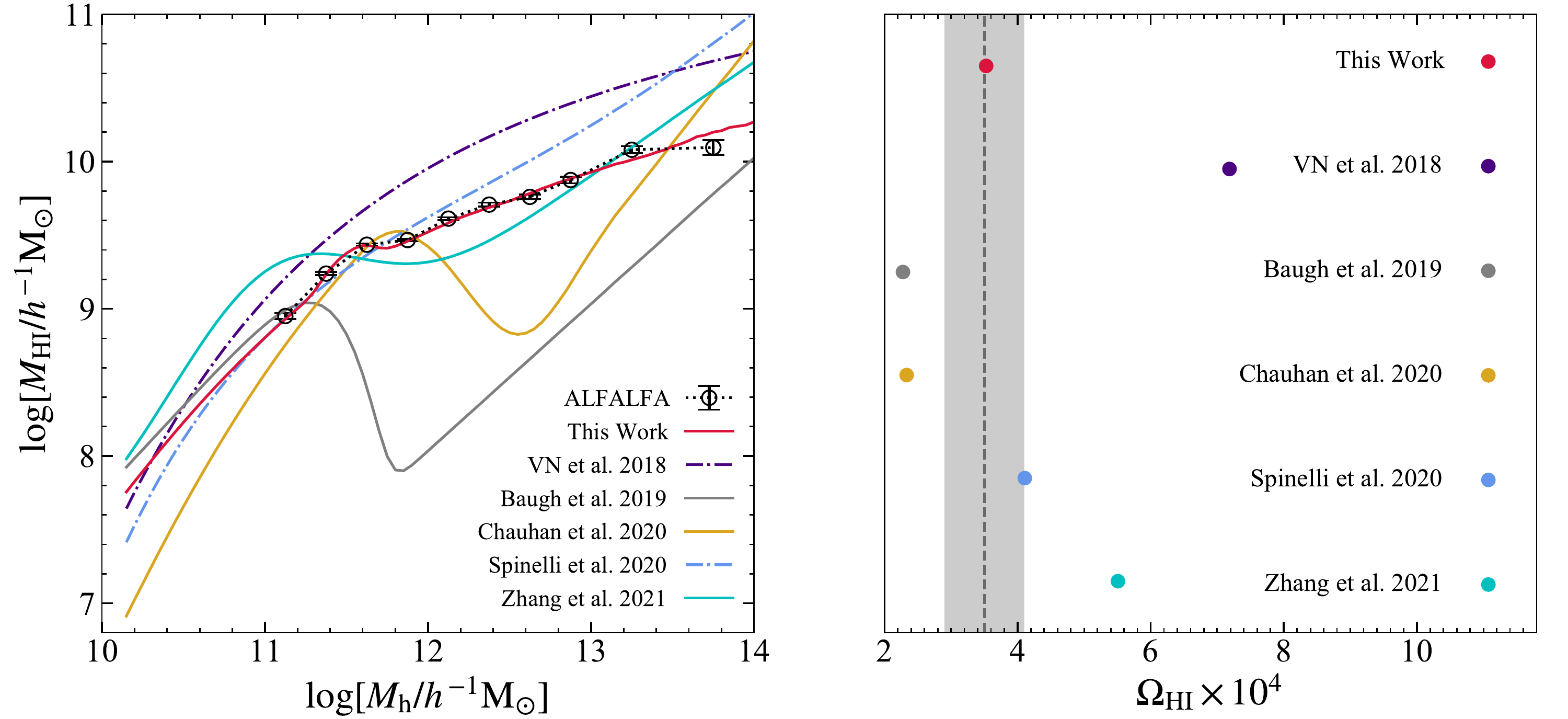}
	\caption{Comparison of different predictions from literature and our best-fitting model. (Left) the average \hi-halo mass relation $\langle M_{\rm HI}|M_{\rm h}\rangle$. We show the semi-analytical models from GALFORM \citep{Baugh2019} (grey solid line), SHARK \citep{Chauhan2020} (yellow solid line), GAEA \citep{Spinelli2020} (blue dot-dashed line), L-GALAXIES \citep{Zhang2021} (cyan solid line), and the analytical fitting formula for TNG100-1 \citep{Villaescusa2018} (purple dot-dashed line), as well as our work (red solid line). We also show the ALFALFA observation result \citepalias{Guo2020} (black circles with error bars). 
	(Right) the \hi\ abundance $\Omega_{\rm HI}$ at $z=0$ for the same models. We also show the observation result (grey dashed vertical line, with light grey shaded area indicating the 1$\sigma$ error) from \cite{Jones2018}. }
	\label{fig:HI-halo}
\end{figure*}

Regarding the comparison between the TNG300-1 and ALFALFA results, the middle and right panels of Figure~\ref{fig:result} show that the \hi\ clustering, especially the projected correlation function $w_{\rm p}(r_{\rm p})$, is significantly larger at large scales in TNG300-1 than in the ALFALFA observation. In other words, the TNG simulation over-predicts the \hi\ bias. 

We also apply the ``null''-model to the halo catalog of the TNG300-1 simulation with the same values of bestfit model parameters as in the ``null''-model obtained by fitting to the TNG100-1 simulation in the left panel of Figure~\ref{fig:result}, and obtain the predictions of \hi\ clustering in the middle and right panels of Figure~\ref{fig:result}. The predicted \hi\ clustering from the ``null''-model agrees well with the direct \hi\ simulation results of TNG300-1. However, this does not imply that the \hi\ mass has no secondary halo dependence in TNG. Instead, this simply implies that in TNG there is weak secondary dependence of \hi\ mass on $c_{\rm vir}$ and $a_{1/2}$, but the dependence on $\lambda$ is much stronger, which is consistent with our findings for TNG in Figure~\ref{fig:SD}. This is because the \hi\ clustering is sensitive to both $c_{\rm vir}$ and $a_{1/2}$, and insensitive to $\lambda$, as shown by comparing the clustering between the ``null''-model and the other secondary dependence models in Figure~\ref{fig:obs}. The fact that the \hi\ clustering is insensitive to halo spin can also be explained as follows. The \hi\ clustering is mostly contributed by halos of $10^{11}\msunh<M_{\rm h}<10^{12}\msunh$, because the abundance of massive halos significantly decreases. However, the dependence of halo clustering (or {\it halo bias}) on $\lambda$ is generally weak for halos in this mass range \citep[see, e.g. Figure~4 of][]{Sato2019}. As the result, the \hi\ clustering depends rather weakly on $\lambda$. 

Note that there are differences in the redshift-space measurements $\xi(s)$ between the ``null''-model and TNG300-1 results at small scale $s<3\mpchi$. This is likely due to the low resolution in the TNG300-1 simulation, since the ``null''-model uses the parameter values obtained by fitting to TNG100-1.

In sum, the TNG simulations show the strong dependence of \hi\ mass on $\lambda$ and weak dependence on $c_{\rm vir}$ and $a_{1/2}$, which is inconsistent with the findings in observations. Also, the \hi\ clustering from the TNG simulations is  significantly over-predicted at large scales, compared to the observations.

\subsection{Comparison with Previous Work}

The \hi-halo mass relation has been recently studied extensively with different galaxy formation models \citep[see,  e.g.][]{Bagla2010,Villaescusa2018,Baugh2019,Spinelli2020,Chauhan2020,Spina2021,Zhang2021}. We compare the results of our empirical model with the different models presented in literature. These models include the semi-analytical models of GALFORM \citep{Baugh2019}, SHARK \citep{Chauhan2020}, GAEA \citep{Spinelli2020}, L-GALAXIES \citep{Zhang2021} and the analytical fitting formula for TNG100-1 \citep{Villaescusa2018}. For a fair comparison with the observation, we apply these different theoretical models to the SMDPL simulation using Eq.~(\ref{eq:avg}), whereby the halo mass binning effect of the observed \hi-halo mass relation is well taken into account. 

We compare the predictions of the \hi-halo mass relation in the left panel of Figure~\ref{fig:HI-halo}. The \hi\ mass in the models of \cite{Villaescusa2018} and \cite{Spinelli2020} increases monotonically with the halo mass, while the \hi-halo mass relation in other models has a non-monotonic feature. As discussed in \citetalias{Guo2020}, this non-monotonic feature is physically related to the virial shock-heating \citep{Rees1977,Silk1977,White1978} and AGN feedback \citep[see e.g.,][]{Baugh2019,Chauhan2020}. It is thus a necessary component of our empirical model. However, the decreasing of $M_{\rm HI}$ in the models of \cite{Baugh2019} and \cite{Chauhan2020} is too strong compared with the observation, while the monotonically increasing models of \cite{Villaescusa2018} and \cite{Spinelli2020} would over-predict $M_{\rm HI}$ in the massive halos. 

It is interesting to note that all of these different models have roughly consistent slopes at the low-mass end, with $M_{\rm HI}$ quickly increasing with $M_{\rm h}$. That is consistent with the three-stage scenario proposed in \citetalias{Guo2020}, i.e.\ the \hi-halo mass relations are dominated by the cold accretion mode in the halo mass ranges of $M_{\rm h}<10^{11.8}\msunh$, by the hot accretion mode in the range of $10^{11.8}<M_{\rm h}<10^{13}\msunh$, and by the mergers in the range of $M_{\rm h}>10^{13}\msunh$, respectively. Figure~\ref{fig:HI-halo} suggests that the \hi\ mass measurements in the halo mass range of $10^{11}-10^{13}\msunh$ can effectively distinguish these different theoretical models, as the cold gas content is very sensitive to the AGN feedback activities \citep{Guo2022}. 

We note that the model of \cite{Chauhan2020} also includes the additional dependence of $M_{\rm HI}$ on the halo spin parameter. The scatter of \hi-halo mass relation in the SHARK semi-analytic model \citep{Lagos2018} is shown to be strongly correlated with $\lambda$, as in the case of TNG. As noted above, including the dependence on $\lambda$ can not resolve the tension of \hi\ bias with observations. 

We show the predicted values of cosmic \hi\ abundance \ $\Omega_{\rm HI}$ at $z=0$ in different models in the right panel of Figure~\ref{fig:HI-halo}. While most of these predictions show large offsets from the ALFALFA observation, our prediction matches best with the observational data. Note that the prediction of the GAEA model \citep{Spinelli2020} shows the best agreement with our prediction, which is due to the similar \hi-halo mass relations at $M_{\rm h}<10^{12}\msunh$ that dominates the contribution to $\Omega_{\rm HI}$. 

These comparisons suggest that our flexible, empirical model works well to accurately describe the \hi\ distribution in the universe. 

\section{Conclusions and Discussions}
\label{sec:conclusion}

In this paper, we propose a flexible, empirical model for the \hi-halo mass relation (see Eqs.~\ref{eq:f_Mh}--\ref{eq:mhi} and Table~\ref{tab:coeff}) that successfully fits the observed cosmic \hi\ abundance ($\Omega_{\rm HI}$), the average \hi-halo mass relation $\langle M_{\rm HI}|M_{\rm h}\rangle$ and the \hi\ clustering measurements for the ALFALFA sample in the redshift range of $0.0025<z<0.06$. 

Our model consists of the primary dependence of \hi\ mass on the mass of host halo and the secondary dependence on other halo parameters, including halo spin parameter ($\lambda$), halo concentration ($c_{\rm vir}$), and halo formation time ($a_{1/2}$). We find that the observation data rejects the ``null''-model, which has no secondary parameters, at $6.49\sigma$ confidence level, and the $\lambda$-model, which has secondary dependence on halo spin parameter, at $5\sigma$ confidence level. We also find that the observed data strongly favors the $a_{1/2}$-model, which has the secondary dependence of \hi\ mass on halo formation time, with a best-fitting $\chi^2/{\rm dof}=27.23/15$. Moreover, the $c_{\rm vir}$-model, which has secondary dependence on halo concentration, is also in agreement with the observation data (but the agreement is slightly worse than the $a_{1/2}$-model). 

In attempt to explain these findings from the perspective of hydrodynamical simulations, we also investigate the \hi-halo mass relation in the TNG simulation, and confirm that the total \hi\ mass in halos shows strong dependence on the halo secondary parameters. However, the TNG simulation favors strong dependence of $M_{\rm HI}$ on $\lambda$, and weak dependence on $c_{\rm vir}$ and $a_{1/2}$. It also predicts the \hi\ clustering with much larger values on large scales than the observations. These results of TNG are in tension with the findings in observations. 

This discrepancy between the TNG simulation and the observation is related to the baryonic physics involved in the cold gas accretion and depletion adopted in the galaxy formation models. It demonstrates the importance of properly modeling the \hi-halo mass relation in order to fully understand the \hi\ distribution in the universe. Comparing our best-fitting model to the previous \hi-halo mass relations in literature, we find substantial difference in both shape and amplitude, which reflects the large uncertainties in the modeling of \hi\ gas. On the other hand, improving the accuracy of measurement and removing systematics of observations might also be necessary to understand the \hi\ distribution. 

An accurate \hi-halo mass relation is of essential applications to construction of mock catalogs for future \hi\ galaxy surveys as well as for the 21~cm intensity mapping surveys. While our model is built on measurements at the local universe, it has the potential to extend to study the evolution of the \hi\ content at higher redshifts, which we leave to explore in future work.

\section*{Acknowledgements}
This work is supported by National SKA Program of China (grant Nos.~2020SKA0110100, 2020SKA0110401), National Key R\&D Program of China (grant Nos.~2018YFA0404502, 2018YFA0404503), NSFC (Nos. 11922305, 11833005, 11821303, 12011530159), and the science research grants from the China Manned Space Project with NO. CMS-CSST-2021-A02. We acknowledge the use of the High Performance Computing Resource in the Core Facility for Advanced Research Computing at the Shanghai Astronomical Observatory.

\software{Multinest \citep{Feroz2009}, CORRFUNC \citep{Sinha2019}, ROCKSTAR \citep{Behroozi2013}, Numpy \citep{Harris_2020}, h5py \citep{collette_python_hdf5_2014}, Matplotlib \citep{Hunter:2007}, Fortran, Python3}

\bibliographystyle{aasjournal}
\bibliography{ref.bib}

\end{document}